\documentclass[preprint,aps]{revtex4}

\usepackage{graphicx}
\usepackage{dcolumn}
\usepackage{bm}

\begin{document}


\title{Frictional magnetodrag between spatially separated 
two-dimensional electron systems:\\
Coulomb versus phonon mediated electron-electron interaction}

\author{Samvel M. Badalyan}%
\email{badalyan@lx2.yerphi.am}
\affiliation{Department of Physics, Chonnam National University, 
Kwangju 500-757, Korea}
\altaffiliation[Permanent address: ]{Radiophysics Department, Yerevan State University, Yerevan, 375025 Armenia.}

\author{Chang Sub Kim}%
\email{cskim@boltzmann.chonnam.ac.kr}
\affiliation{Department of Physics, Chonnam National University, Kwangju 500-757, Korea}%

\date{\today}

\begin{abstract}

We study the frictional drag due to Coulomb and phonon mediated
electron-electron interaction in a double layer electron system exposed to a
perpendicular magnetic field. Within the random phase approximation we 
calculate the dispersion relation of the intra Landau level magnetoplasmons 
at finite temperatures and distinguish their contribution to the magnetodrag.
We calculate the transresistivity $\rho _{Drag}$ as a function of magnetic 
field $B$, temperature $T$, and interlayer spacing $\Lambda $ for a matched
electron density.
For $\Lambda =200$ nm we find that $\rho _{Drag}$ is solely due to phonon 
exchange and shows no double-peak structure as a function of $B$. 
For $\Lambda =30$ nm, $\rho _{Drag}$ shows the double-peak structure and is
mainly due to Coulomb interaction.
The value of $\rho _{Drag}$ is about $0.3$ $\Omega $ at $T=2$ K and for the 
half-filled second Lanadau level, which is about $13$ times larger than 
the value for $\Lambda =200$ nm. 
At lower edge of the temperature interval from 
$0.1$ to $8$ K, $\rho _{Drag}/ T^{2}$ remains finite for $\Lambda =30$ nm 
while it tends to zero for $\Lambda =200$ nm. Near the upper edge of this 
interval, $\rho_{Drag}$ for $\Lambda =30$ nm is approximately linear in 
$T$ while for $\Lambda =200$ nm it decreases slowly in $T$. 
Therefore, the peak of $\rho_{Drag}/ T^{2}$ is very sharp for 
$\Lambda =200$ nm. This strikingly different magnetic field and 
temperature dependence of $\rho _{Drag}$ ascribe we mainly to the 
weak screening effect at large interlayer separations.

\end{abstract}

\pacs{Valid PACS appear here}
\maketitle

\section{Introduction}

Frictional drag between two spatially separated two-dimensional (2D) electron
systems is a powerful tool to probe electron-electron ($e-e$) interaction
and is of current intensive research interest both experimentally and
theoretically \cite{eisen,rojo}. The drag effect manifests itself when a
current with a density $J_{1}$ driven along the layer $1$ induces, via
momentum transfer, an electric field $E_{2}$ in the layer $2$ under the
condition that the layer $2$ is an open circuit \cite{pogreb,price}.

Much experimental work was done in the magnetic field-free case to measure
the drag effect in a coupled 2D and 3D electron system \cite{solomon},
between two 2D electron systems (2DES) \cite
{gramila91,gramila93,rubel95,hill97,noh98,noh99,kellogg}, two 2D hole
systems (2DHS) \cite{joerger,pillarisetty}, and also in a coupled 2DES-2DHS 
\cite{sivan,joerger}. Theoretical work on the frictional drag has been
devoted to explain different momentum transfer mechanisms in a coupled
2DES-3DES \cite{laikhtman,boiko92}, 2DES-2DHS \cite
{tso93,swierkowski95,vignale}, 2DES-2DES \cite
{boiko90,maslov,tso92,zhang,jauho,zheng,kamenev,flensberg94,flensberg95,flensberg955,swierkowski97,bons98,smb}%
, and recently in a 2DHS-2DHS \cite{hwang}. Also, plasmon-enhancement of the
drag due to many-body correlations at high temperatures has been studied
theoretically \cite{flensberg94,flensberg955} and experimentally \cite
{hill97,noh98}. Two possible interlayer $e-e$ interaction mechanisms have
been considered: One is the direct electrostatic Coulomb scattering and the
other is the phonon-mediated effective interaction. The Coulomb drag
calculations predict a $T^{2}$ temperature dependence for the
transresistivity $\rho _{_{Drag}}$ and a strong inter-layer spacing ($%
\Lambda )$ dependence as $\Lambda ^{-4}$ \cite{tso92,jauho}$.$ However, the
experimental results do not support the pure Coulomb mechanism \cite
{gramila93,rubel95,noh99}. The observed drag has a peak when the electron
densities match, $n_{1}=n_{2}$, and the scaled transresistivity, $%
\rho_{_{Drag}}/T^{2}$, deviates from a constant value demonstrating a peak
as a function of $T$ well below Bloch-Gr\"{u}neisen temperature. Also, the
measured $\rho _{_{Drag}}$ for separations at least up to 50 nm exhibits
almost no dependence on $\Lambda $ and its value beyond $\Lambda =50$ nm is
much bigger to be accounted for by Coulomb interaction alone. This led
Gramila \textit{et al}. \cite{gramila93} to propose an acoustical
phonon-mediated scattering mechanism for the drag. This mechanism was
studied extensively in Refs.~\onlinecite{tso92,zhang,bons98,smb}. The
theoretical results are satisfactory to explain the main experimental
findings. The phonon-mediated drag due to the bare electron-phonon ($e-p$)
coupling diverges when the transferred energy, $\hbar \omega ,$ from the
layer $1$ to the layer $2$ is close to $\hbar sq$ ($\overrightarrow{q}$ is
the transferred in-plane momentum). This divergence is eliminated by taking
into account other scattering mechanisms such as screening and/or the finite
phonon mean free path \cite{bons98}. It is found that, despite the weak $e-p$
coupling, the effective phonon-mediated $e-e$ interaction remains strong and
competes with the Coulomb interaction. In general, $e-p$ scattering in 2DESs
is qualitatively different between in the small angle scattering (SAS)
temperature region, $T\ll \hbar sk_{F}$, and in the large angle scattering
(LAS) temperature region, $T\sim \hbar sk_{F}$, where $s$ is the speed of
sound and $k_{F}$ the Fermi wave vector \cite{karpus,smb2}. In the SAS
region, $\hbar \omega \lesssim T$ and this results in the power law
dependence of $\rho _{Drag}\propto T^{6}.$ On the other hand, in the LAS
region, $\hbar \omega \approx 2\hbar sk_{F}$ and $\rho _{Drag}$ varies
linearly in $T$. As a consequence, $\rho _{Drag}/T^{2}$ has a peak at the
crossover from the SAS to LAS temperature region at the peak temperature $%
T_{peak}<\hbar sk_{F}$. In the LAS temperature region the momentum transfer
is most efficient when momenta of the emitted and absorbed phonons are
determined by the same Fermi wave vector $k_{1F}=k_{2F}$, this results in
the peaked transresistivity when $n_{1}=n_{2}$. In the SAS region there is
no cutoff related to $k_{F}$ that $\rho _{Drag}$ shows a monotonous behavior
as a function of $n_{2}$.

The frictional drag has been also investigated in the presence of a
perpendicular magnetic field. The drag rate has been measured experimentally
in a coupled 2DES \cite
{hill96,hill97,rubel97,rubel97e,patel,lilly,feng,joerger2000e}, 2DHS \cite
{joerger} as well in a 2DES-2DHS \cite{rubel97e}. Also, the response of
composite fermions to large wave vector scattering has been studied through
phonon magnetodrag measurements \cite{zelakiewicz}. In addition, recent
measurements reveal a new regime of the magnetodrag in a coupled 2DES at
mismatched densities $n_{1}\neq n_{2}$ in which the polarity of the drag
voltage is opposite that normally found for electron systems \cite{feng,lok}%
. Theoretical investigations were mostly carried out with incorporating the
direct Coulomb mechanism \cite{bons96,bons97,wu,manolescu}. These
calculations show reasonable agreement with the experiments \cite
{hill96,rubel97} that at low temperatures $\rho _{Drag}$ demonstrates an
analog of the Shubnikov-De Haas oscillations with amplitudes by two order of
magnitude enhanced compared to the zero field drag signal. At the matched
densities the measured magnetodrag has a double-peak structure in magnetic
field in the inter-quantum Hall plateau regions \cite{rubel97,feng}. This
observation is in agreement with the earlier calculations of the Coulomb
magnetodrag by B\o nsager \textit{et al}. \cite{bons96,bons97} who predicted
this behavior caused by the interplay of screening and Landau quantization.
However, the calculations by Wu \textit{et al.} \cite{wu} and the
experiments by Hill \textit{et al.} \cite{hill96}\ show oscillations without
the double peak structure. The critical test of the theory was performed in
a recent experiment by Lok \textit{et al.} \cite{lok}, which confirms that
the transresistivity does not show the predicted double peak structure for
the spin split Landau levels and the double-peak structure at higher filling
factors is not caused by the screening effect \cite{joerger2000e}. Notice
also, as against to the zero magnetic field case, the scaled
transresistivity $\rho _{Drag}/T^{2}$ due to the pure Coulomb mechanism
shows a peaked temperature dependence in the finite magnetic fields \cite
{bons96,bons97}. In addition, the Coulomb magnetodrag calculations \cite
{khaetskii,wu,manolescu} reveal new features associated with interlayer
magnetoplasmons, that was observed\ also experimentally \cite{hill97}.

As we described in the previous paragraph, the phonon-mediated drag is the
dominant mechanism in the zero magnetic field case in a coupled 2DES, except
for very closely spaced layers \cite
{gramila93,rubel95,noh99,joerger,bons98,smb}. Further, it was reported
recently that the effective phonon and the Coulomb contributions to drag
were comparable in a coupled 2DHS even with closely spaced layers. The
phonon-mediated drag effect has been also investigated experimentally in the
finite magnetic fields \cite{rubel97e,joerger} where it was found also that
the phonon contribution was prevalent in the coupled 2DES, 2DHS, and
2DES-2DHS with large barriers. Up to now, however, it is rare to find
theoretical works on the frictional drag mediated by phonon exchange in a
finite magnetic field. The exceptions are recent treatments of the phonon
drag at the Landau level filling factor $\nu =1/2$ by Chern-Simons composite
fermion theory \cite{khveshchenko,bons2000}.

The aim of the present paper is to investigate the direct Coulomb and the
effective phonon-mediated frictional drag in a coupled 2DES exposed to a
perpendicular magnetic field theoretically. For this system we calculate the
dispersion relation of the intra-Landau level magnetoplasmons by taking into
account finite temperature and distinguish the magnetoplasmon contribution
to the magnetodrag. We calculate $\rho _{Drag}$ as a function of magnetic
field $B$, temperature $T$, and interlayer spacing $\Lambda $, and determine
the relative contributions of Coulomb versus phonon-mediated $e-e$
interaction to magnetodrag

This paper is organized as follows. In the next Sec. II we describe briefly
the theoretical method we use to calculate the transresistivity, which
follows mainly to Refs.~\onlinecite{bons97,bons98}. In Sec. III we present
and discuss our calculations. Finally, the results are summarized and the
conclusions are given in Sec. IV.

\section{Theory}

\begin{figure}[tbp]
\centering
\includegraphics[width=5cm]{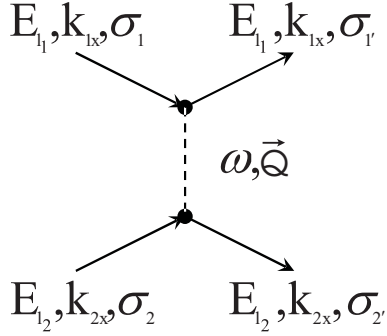}
\caption{Diagram for the effective phonon mediated $e-e$ interaction. Here $%
E_{l}$, $k_{x}$, and $\protect\sigma $ are the electron energy, the momentum
in the Landau gauge, and the spin, respectively.}
\label{fg1}
\end{figure}

Our theoretical model consists of two parallel electron layers exposed to a
perpendicular magnetic field in the $z$-direction, separated by a distance $%
\Lambda $ from center to center, each having the electron densities $n_{1}$
and $n_{2}$ and the layer extensions $d_{1}$ and $d_{2}$. The experimentally
measured quantity is the transresistivity defined as $\rho _{Drag}={%
E_{2}/J_{1}}$. The main theoretical framework to calculate $\rho _{Drag}$ is
the Fermi liquid theory where the \ interlayer $e-e$ interaction is treated
perturbatively. The expression for $\rho _{Drag}$ can be derived by using
the linearized Boltzmann equation \cite{jauho} or the memory function
formalisms starting from the Kubo formula \cite
{zheng,kamenev,flensberg95,bons98}. Direct calculations show that in our
model the transresistivity is represented in the standard form 
\begin{eqnarray}
\rho _{Drag} &=&-{\frac{\hbar ^{2}}{2e^{2}n_{1}n_{2}T}\frac{1}{A}}\sum_{%
\overrightarrow{q}}q^{2}\int_{0}^{\infty }\frac{d\omega }{2\pi }\frac{\left|
W_{12}(q,\omega )\right| ^{2}}{\left| \varepsilon (q,\omega )\right| ^{2}} 
\nonumber \\
&&\times \frac{\text{Im}\chi _{1}(q,\omega )\text{Im}\chi _{2}(q,\omega )}{%
\sinh ^{2}(\hbar \omega /2T)}  \label{eq2}
\end{eqnarray}
where $A$ is the normalization area, $W_{ij}(q,\omega )=\sum_{\Upsilon
}W_{ij}^{\Upsilon }(q,\omega )$ the total unscreened interlayer interaction
matrix element ($i,j=1,2$ are the layer indices), $\varepsilon (q,\omega )$
the screening function, and $\chi _{i}(q,\omega )$ the irreducible
intralayer electron polarization functions (the interlayer electron
polarization is neglected). The index $\Upsilon $ refers to the type of $e-e$
interaction. We consider direct Coulomb ($\Upsilon =$ C) $e-e$ interaction
and effective $e-e$ interaction, mediated by the exchange of piezoelectric ($%
\Upsilon =$ PA) and deformation ($\Upsilon =$ DA) acoustical phonons as well
as the exchange of polar optical phonons ($\Upsilon =$ PO). Phonon mediated $%
e-e$ interaction appears in second order perturbation theory with respect to
the bare $e-p$ coupling. The unscreened interaction matrix element of this
process is visualized by the diagram in Fig.~\ref{fg1}. The free phonon
propagator 
\begin{equation}
D^{\Upsilon }(Q,\omega )={2\omega _{Q}^{\Upsilon }\hbar ^{-1}\left[ \left(
\omega +\frac{i}{2\tau _{Q}^{\Upsilon }}\right) ^{2}-\left( \omega
_{Q}^{\Upsilon }\right) ^{2}\right] }^{{-1}}  \label{eq3}
\end{equation}
with $\overrightarrow{Q}=(\overrightarrow{q},q_{z})$, $\omega
_{Q}^{PA,DA}=sQ $ and $\omega _{Q}^{PO}=\omega _{_{PO}}$, and a summation
over phonon momenta corresponds to the internal phonon line (the dashed line
in the diagram). $\tau _{Q}^{\Upsilon }$ is the lifetime of the phonons. The
vertex part of this diagram (the solid dots in Fig.~\ref{fg1}) corresponds
to 
\begin{equation}
\Gamma _{i}^{\Upsilon }=\sqrt{B^{\Upsilon }(Q)}I_{i}(q_{z}d)
\end{equation}
where $B^{\Upsilon }(Q)$ are the bare $e-p$ coupling functions \cite{gantlev}
and $I_{i}(\xi )=\int dz\rho _{i}(z)e^{i\xi z/d}$ are the form factors in $z$%
-direction (the remaining in-plane part of the form factor enters the
definition of $\chi (q,\omega )$). We assume that electron scattering takes
place in the lowest electron subbands of infinitely high quantum wells with
widths$\ d_{1}=d_{2}=d$ and the subband density functions $\rho _{2}(z)=\rho
_{1}(z+\Lambda )$ in both layers do not depend on the subband index. 
In GaAs the PA interaction matrix element is given by the effective
piezoelectric modulus which depends only on the polarization of the phonon
and its direction of propagation. Since in this description of frictional
drag we are interested in the average relaxation characteristics of the
system, the anisotropy of PA interaction can be ignored. We describe PA
interaction within the scope of the isotropic model \cite{gantlev} that
leads to a scalar PA interaction constant. We assume also that all elastic
parameters of the sample are the same and phonons are not reflected from the
interfaces separating different materials (for instance between GaAs and
AlGaAs).

Taking the summation over $q_{z}$ and including also the Coulomb
contribution, we obtain for the total unscreened interlayer $e-e$
interaction matrix elements 
\begin{eqnarray}
W_{ij}(q,\omega ) &=&\frac{2\pi e^{2}}{\kappa _{0}}\frac{1}{q}\left\{
F_{ij}(qd)+\frac{\overline{\tau }_{C}}{\overline{\tau }_{PO}}\frac{\omega
_{PO}^{2}}{\omega _{PO}^{2}-\omega ^{2}}F_{ij}(qd)\right.  \nonumber \\
&&+\frac{\overline{\tau }_{C}}{\overline{\tau }_{PA}}\frac{q}{\alpha }%
F_{ij}(\alpha d)  \nonumber  \label{eq8} \\
&&\left. +\frac{\overline{\tau }_{C}}{\overline{\tau }_{DA}}\left(
F_{3}\delta _{ij}+\frac{\omega ^{2}}{2\left( sp_{_{PO}}\right) ^{2}}\frac{q}{%
\alpha }F_{ij}(\alpha d)\right) \right\}
\end{eqnarray}
where $\kappa _{0}=13.1$ \cite{adachi} is the GaAs static dielectric
constant, $\hbar p_{_{PO}}=\sqrt{2m\hbar \omega _{_{PO}}},$ $\omega _{_{PO}}$
the polar optical phonon frequency, and $\alpha =\sqrt{q^{2}-(\frac{\omega }{%
s}+\frac{i}{2s\tau _{q}^{\Upsilon }})^{2}}$. In the limit of infinite phonon
lifetime $\alpha =\sqrt{q_{\perp }^{2}-\omega ^{2}/s^{2}}$ if $q_{\perp
}^{2}>\omega ^{2}/s^{2}$ and $\alpha =-i\sqrt{q_{\perp }^{2}-\omega
^{2}/s^{2}}$ if $q_{\perp }^{2}<\omega ^{2}/s^{2}$ with the positive square
root branch. The form factors $F$ are given by 
\begin{eqnarray}
F_{ij}(\xi ) &=&\left\{ 
\begin{array}{c}
e^{-\frac{\Lambda }{d}\xi }F_{1}(\xi ),\text{if } i\neq j \\ 
F_{2}(\xi ),\text{if } i=j,
\end{array}
\right. ,F_{1}(\xi )=I(\xi )I(-\xi ),  \nonumber  \label{eq9} \\
F_{2}(\xi ) &=&\int dz_{1}dz_{2}\rho (z_{1})\rho (z_{2})e^{i\xi \frac{\left|
z_{1}-z_{2}\right| }{d}},F_{3}=d\int dz\rho ^{2}(z).
\end{eqnarray}
The following nominal scattering times are introduced in Eq.~(\ref{eq8}) 
\begin{eqnarray}
\frac{1}{\overline{\tau }_{_{C}}} &=&\frac{2\omega _{PO}}{p_{_{PO}}a^{\ast }}%
,\text{\ }\frac{1}{\overline{\tau }_{_{PA}}}=\frac{B_{0}^{PA}p_{_{PO}}}{2\pi
\hbar ^{2}s}{,}\text{\ }B_{0}^{PA}={\frac{\hbar (e\beta )^{2}}{2\varrho s}},
\nonumber \\
\frac{1}{\overline{\tau }_{_{PO}}} &=&{2\alpha _{F}\omega }_{{PO}}{,}\text{\ 
}\frac{1}{\overline{\tau }_{_{DA}}}=\frac{B_{0}^{DA}p_{_{PO}}^{3}}{\pi \hbar
^{2}s},\text{\ }B_{0}^{DA}={\frac{\hbar \Xi ^{2}}{2\varrho s}}
\end{eqnarray}
where ${\alpha _{F}}$ is the Fr\"{o}lich interaction constant, $\Xi $ and $%
e\beta $ the deformation and piezoelectric phonon potential constants, $%
\varrho $ the crystal mass density, and $a^{\ast }$ the effective Bohr
radius. For GaAs we have $\overline{\tau }_{C}\approx 0.024$ ps and the
numerical values $\overline{\tau }_{PO}\approx 0.14$ ps, $\overline{\tau }%
_{PA}\approx 8$ ps, $\overline{\tau }_{DA}\approx 4$ ps take we from \cite
{gantlev}.

We are interested in the experimental situations when the Landau levels are
fully resolved, accordingly the scale of $\omega $ over which a 2DES
responds to an external perturbation is given by the width of the Landau
band. Therefore, when $\omega \ll \omega _{PO}$, the optical phonon
propagator $D^{PO}(q,\omega )$ is never on the mass surface, and only
virtual optical phonons can contribute to the magnetodrag in this regime.
The PO phonon mediated contribution to total interaction, $%
W_{ij}^{PO}(q,\omega )$ (the second term in Eq.~(\ref{eq8})), at $\omega =0$%
, behaves like the Coulomb contribution, $W_{ij}^{C}(q,\omega )$ (the first
term in Eq.~(\ref{eq8})). And, because of the small $\overline{\tau }_{C}/%
\overline{\tau }_{PO}$ coupling, simply results in about $17\%$
renormalization of $W_{ij}^{C}(q,\omega )$. From Eqs.~(\ref{eq8}) and (\ref
{eq9}) one can see that $W_{12}^{C}(q,\omega )$ is large when $q\Lambda
\lesssim 1$. In drag experiments $\Lambda $ is larger than $d$ so we have
usually $qd\ll 1.$ The PA and DA phonon contributions, $W_{ij}^{PA,DA}(q,%
\omega )$ (the third and forth terms in Eq.~(\ref{eq8})), behave like a
delta function because of the small phonon couplings $\overline{\tau }_{C}/%
\overline{\tau }_{PA}\approx 3\cdot 10^{-3}$ and $\overline{\tau }_{C}/%
\overline{\tau }_{PA}\approx 1.5\cdot 10^{-3}$: $W_{ij}^{PA,DA}$ are
negligible with respect to the Coulomb term $W_{ij}^{C}(q,\omega )$ for all
over $\omega $ and $q$ except the close neighborhood of $\omega =sq$ where $%
\alpha $ is very small and we have again $\alpha d\ll 1$. Thus, the
zero-limit approximation of the form factors $F$ is well justified for both
Coulomb and phonon mediated $e-e$ interaction and we take $F_{1,2}(\xi
)\approx F_{1,2}(0)$ in our numerical calculations.

In the random phase approximation the dielectric tensor is obtained from the
solution of a matrix Dyson equation for the dynamically screened interlayer
interaction \cite{sarma} and has the form $\varepsilon _{ij}(q,\omega
)=\delta _{ij}-W_{ij}(q,\omega )\chi _{j}(q,\omega )$ (assuming no summation
by the repeating indices). The screening function $\varepsilon (q,\omega )$
is the determinant of the dielectric tensor and is given by 
\begin{eqnarray}
\varepsilon (q,\omega ) &=&\left( 1-W_{11}(q,\omega )\chi _{1}(q,\omega
)\right) \left( 1-W_{22}(q,\omega )\chi _{1}(q,\omega )\right)  \nonumber \\
&&-W_{12}^{2}(q,\omega )\chi _{1}(q,\omega )\chi _{2}(q,\omega )
\end{eqnarray}
In this approximation $\chi (q,\omega )$ is determined by the bubble
diagrams with the exact vertex part. B\o nsager \textit{et al.} \cite{bons97}
have shown that when the Landau levels are clearly resolved, corrections to
the vertex part are small. In the first approximation they can be neglected
and $\chi (q,\omega )$ is given by 
\begin{eqnarray}
\chi (q,\omega ) &=&\frac{1}{\pi \ell _{B}^{2}}\sum_{l,l^{\prime
}=0}^{\infty }Q_{ll^{\prime }}^{2}\left( \frac{q^{2}\ell _{B}^{2}}{2}\right)
\int_{0}^{\infty }\frac{dE}{\pi }f_{T}(E-E_{F})  \nonumber \\
&&\times \text{Im}G_{l}^{R}(E)\left( G_{l^{\prime }}^{R}(E+\hbar \omega
)+G_{l^{\prime }}^{A}(E-\hbar \omega )\right) ,  \label{eq14}
\end{eqnarray}
$\ell _{B}$ is the magnetic length and $f_{T}$ the Fermi distribution
function determined by the chemical potential $E_{F}.$ The bare vertex
functions $Q_{ll^{\prime }}$ are given by the gauge invariant part of the
in-plane form factor 
\begin{equation}
Q_{ll^{\prime }}\left( t\right) =\left( l!/l^{\prime }!\right)
^{1/2}e^{-t/2}t^{(l-l^{\prime })}L_{l^{\prime }}^{l-l^{\prime }}\left(
t\right) ,\text{ for }l\leq l^{\prime }
\end{equation}
where $L_{l^{\prime }}^{l-l^{\prime }}\left( t\right) $ is the associated
Laguerre polynomial. For $l^{\prime }>l$ one can exploit the symmetry
relation $Q_{ll^{\prime }}\left( t\right) =(-1)^{l-l^{\prime }}Q_{ll^{\prime
}}\left( t\right) .$ In Eq.~(\ref{eq14}) $G_{l}^{R,A}(E)$ are the electron
retarded and advanced Green functions dressed by electron-impurity ($e-i$)
interaction. We treat $e-i$ scattering in the short-range impurity model 
\cite{afs} and use the Green functions obtained by Ando and Uemura \cite
{ando} in the self-consistent Born approximation 
\begin{equation}
G_{l}^{R,A}(E)=2\left[ E-E_{l}+\sqrt{\left( E-E_{l}\right) ^{2}-\Gamma
_{0}^{2}}\right] ^{-1}.
\end{equation}
In this approximation the half-width $\Gamma _{0}=\sqrt{\frac{2}{\pi }\hbar
\omega _{B}\frac{\hbar }{\tau }}$ of the Landau level $E_{l}=(l+1/2)\hbar
\omega _{B}$\ is independent of the Landau index $l$ ($\omega _{B}$ is the
cyclotron frequency and $\tau $ is the transport relaxation time that
determines the current in the drive layer via the mobility $\mu =e\tau
/m^{\ast }$, $m^{\ast }$ is the electron effective mass). The chemical
potential $E_{F}(n,B,T,\mu )$ in distribution functions is determined
implicitly from the electron density 
\begin{equation}
n=-\frac{1}{2\pi \ell _{B}^{2}}\sum_{l=0}^{\infty }\int_{0}^{\infty }\frac{dE%
}{\pi }f_{T}(E-E_{F})\text{Im}G_{l}(E).
\end{equation}
Here we use the Green functions obtained by Gerhardts \cite{gerhardts} in
the improved self-consistent Born approximation 
\begin{eqnarray}
G_{l}(E) &=&-i\frac{\sqrt{2\pi }}{\Gamma _{0}}\exp \left( -2\frac{\left(
E-E_{l}\right) ^{2}}{\Gamma _{0}^{2}}\right)  \nonumber \\
&&\times \left[ 1+Erf\left( i\sqrt{2}\frac{E-E_{l}}{\Gamma _{0}}\right) %
\right]  \label{eq18}
\end{eqnarray}
which correspond to the Gaussian density of states without the unphysical
edges of Landau bands. $Erf$(z) is the error function.

Thus, Eqs.~(\ref{eq2}) and (\ref{eq8}-\ref{eq18}) determine the
transresistivity of the double layer 2DES exposed to a perpendicular
magnetic field and coupled via direct Coulomb and effective phonon mediated $%
e-e$ interaction. In this treatment the Landau levels acquire broadening due
to $e-i$ scattering and each broadened state has its weighted contribution
to drag. The electron energy remains dispersionless and the distribution
functions do not depend on the position of the Landau oscillator center. For
this reason $\rho _{Drag}$ does not depend on the gauge non-invariant
electron momenta $k,$ as it should be, but also after performing the
summation over all $k,$ no interference occurs in Eq.~(\ref{eq2}) between
different Fourier components $\overrightarrow{q}$ both of the Coulomb \cite
{smb93} and the $e-p$ interaction potentials. It should be noticed also that
only such non-trivial treatment of $e-e$, $e-p$, and $e-i$ scattering allows
the existence of the intra-Landau band magnetoplasmons and $e-p$ scattering
processes with a finite energy transfer $\hbar \omega \neq 0$. This is
especially important in the regime of $T$, $\Gamma _{0}\ll $ $\hbar \omega
_{B}$ under the consideration here. Without impurity broadening of the
Landau levels, the phonon mediated magnetodrag in this regime is only due to
exchange of pure virtual phonons with $\hbar \omega =0$ and $q\neq 0$. But
this contribution is small with respect to the Coulomb magnetodrag for all
phonon modes. In contrast to this, for finite $\hbar \omega $, acoustical
phonon mediated $e-e$ interaction diverges in $\omega =sq$. This divergence
enhances strongly the phonon contribution to magnetodrag and makes it
dominant at large interlayer separations.

\section{Result and discussion}

Below we present our calculations of $\rho _{Drag}$ as a function of $%
B$, $T$, and $\Lambda $ in a symmetric double layer quantum well system with
a matched electron density $n=2.5\cdot 10^{15}$ m$^{-2}$. We neglect the
spin splitting and assume that the Landau levels are fully resolved and the
odd filling factors correspond to a half-filled Landau levels. The spin
degeneracy results in an additional factor of 4 in Eq.~(\ref{eq2}). It is 
known that at
temperatures below Bloch-Gr\"{u}neisen temperature the frictional drag is 
dominated by the PA phonons in the case of $B=0$ \cite{smb}, and the PO
contribution is weak as we describe in the previous Section. Here we consider
mainly PA phonon mediated $e-e$ interaction. The inclusion of calculations for
DA phonon mediated $e-e$ interaction is straightforward.
\begin{figure}[tbp]
\centering
\includegraphics[width=10cm,angle=-90]{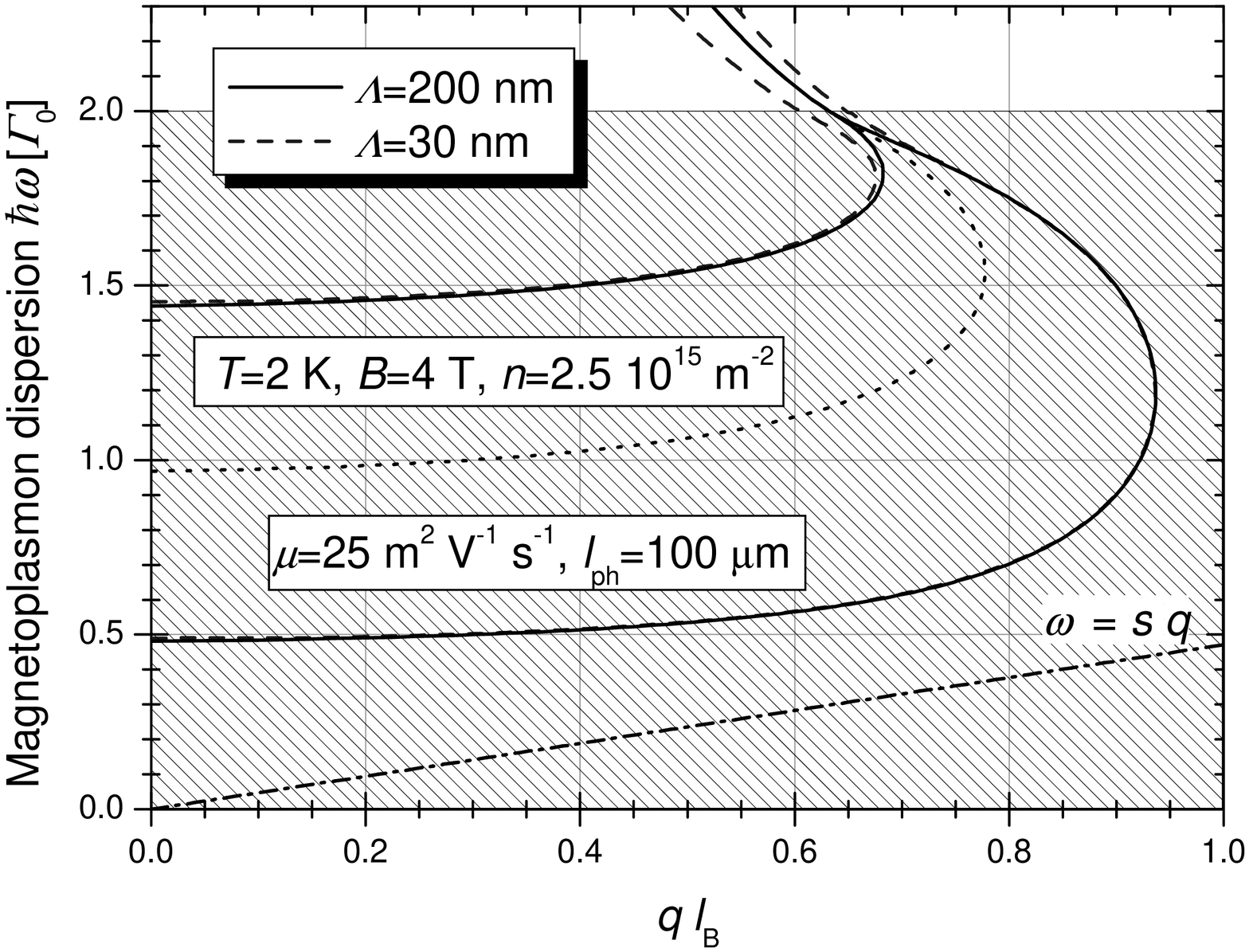}
\caption{Intra Landau level magnetoplasmon dispersion relation for the
adopted parameters obtained from the zeros of the real part of the screening
function. Dotted line shows zeros of the real part of the screening
function for a single layer.}
\label{fg6}
\end{figure}
We take into account only the intra-Landau level
magnetoplasmon modes and, in the low temperature regime ($T\ll \omega _{B}$)
under consideration here, we neglect the interlevel magnetoplasmon
contributions to the magnetodrag. The intra-Landau level collective excitations
\cite{anton} exist when a Landau level is partially filled.
In Fig.~\ref{fg6} we plot the intra-Landau
level magnetoplasmon dispersion relation obtained from the zeros of the real
part of the screening function, Re~$\varepsilon \left( q,\omega \left(
q\right) \right) =0$, with the imaginary part describing the damping. The
spectra have two collective modes with the finite values $\omega ^{-}\left(
0\right) $ and $\omega ^{+}\left( 0\right) $ close to $\Gamma _{0}/2$ and $%
3\Gamma _{0}/2$ in the limit of $q\rightarrow 0$. When $B$ varies in the
second inter-plateau region from $3$ to $4.5$ T for $T=2$ K or $T$ varies
from $1$ to $8$ K for $B=4$ T, $\omega ^{-}\left( 0\right) $ ($\omega
^{+}\left( 0\right) $) is changed slightly, within $1\%$ ($5\%$). The values
of $\omega ^{-}\left( 0\right) $ and $\omega ^{+}\left( 0\right) $ increase
monotonically in $T$ in this temperature interval. In this magnetic field
interval, they approach closely each other at $B$ corresponding to the
half-filled Landau level. The whole spectra are located in a finite range of 
$q$ close to zero so that the magnetoplasmons have momenta $q\ell _{B}<1$
for the choice of parameters in Fig.~\ref{fg6}. For such momenta the phonon
energy $\hbar sq$ (the dashed-dotted line in Fig.~\ref{fg6}) lies below the
magnetoplasmon energy near the lower edge of Landau band. Therefore, the
phonon mediated $e-e$ interaction effect on the magnetoplasmon spectrum is
negligible. For energies $\hbar \omega $ well below $2\Gamma _{0}$ ($\Gamma
_{0}\approx 6.5$ K, $\Gamma _{0}/\hbar \omega _{B}\approx 0.08$ for $B=4$
T), both the splitting and broadening of each magnetoplasmon mode of the
spectra are determined by the strong damping Im~$\chi (q,\omega (q))$. The
poles corresponding to these collective modes are moved into the complex $%
\omega $-plane and are located far from the real axis. Therefore, the two
modes of the spectra are strongly mixed and damped and cannot be interpreted
as the normal modes of the system. 
We note that in this strongly dispersive system, the
group velocity $v_{g}\equiv d\omega (q)/dq$ shows anomalous behavior: It
becomes infinite and changes its sign at certain points. This should not
cause an alarm because in the regions of anomalous dispersion with Re~$%
\varepsilon \left( q,\omega \left( q\right) \right) =0$, the approximation $%
\Delta \omega (q)\approx v_{g}q$ is not valid and the group velocity is not
a useful concept \cite{jackson}. Noticed also that the situation when the
group velocity changes its sign in a finite wavenumber arises also, for
instance, in the single particle spectra of magnetic edge states in 1D \cite
{smbfmp} and 2D \cite{peetmat,fmp} electron systems exposed to a
nonhomogeneous magnetic field. Towards the upper edge of the Landau band the
damping decreases and becomes zero for $\hbar \omega \geq 2\Gamma _{0}$.
Thus, in this model of the Landau band with sharp unphysical edges, the
magnetoplasmons with energies $\hbar \omega \geq 2\Gamma _{0}$ have no
contribution to the magnetodrag. Near the upper edge of Landau band, the
spectra are splitted into the symmetric and antisymmetric magnetoplasmon
modes and this is due to the interlayer coupling, determined by the spacing $%
\Lambda $. One can see that for $\Lambda =30$ and $200$ nm the
magnetoplasmon dispersion shows different behavior in the close vicinity of $%
\hbar \omega =2\Gamma _{0}$. In this region damping is vanishingly small,
and the spectra describes long lifetime bound states of magnetoplasmons in a
double layer system with a binding energy given by the splitting of the
magnetoplasmon modes. For $\Lambda =30$ nm the binding energy is about $1$ K
near the upper edge of the Landau band, and it decreases strongly with $%
\Lambda $. For $\Lambda =200$ nm\ Coulomb interlayer $e-e$\ interaction is
obviously very weak, the binding energy is negligibly small, and the
magnetoplasmon dispersion becomes degenerated. We note that the spectrum of
inter Landau level magnetoplasmons near the upper edge of the Landau band
has been calculated analytically for pure Coulomb $e-e$ interaction by
Khaetskii and Nazarov in Ref.~\onlinecite{khaetskii}, where the analogous
dispersion curves for higher cyclotron harmonics was obtained.

\begin{figure}[tbp]
\centering
\includegraphics[width=6cm,angle=-90]{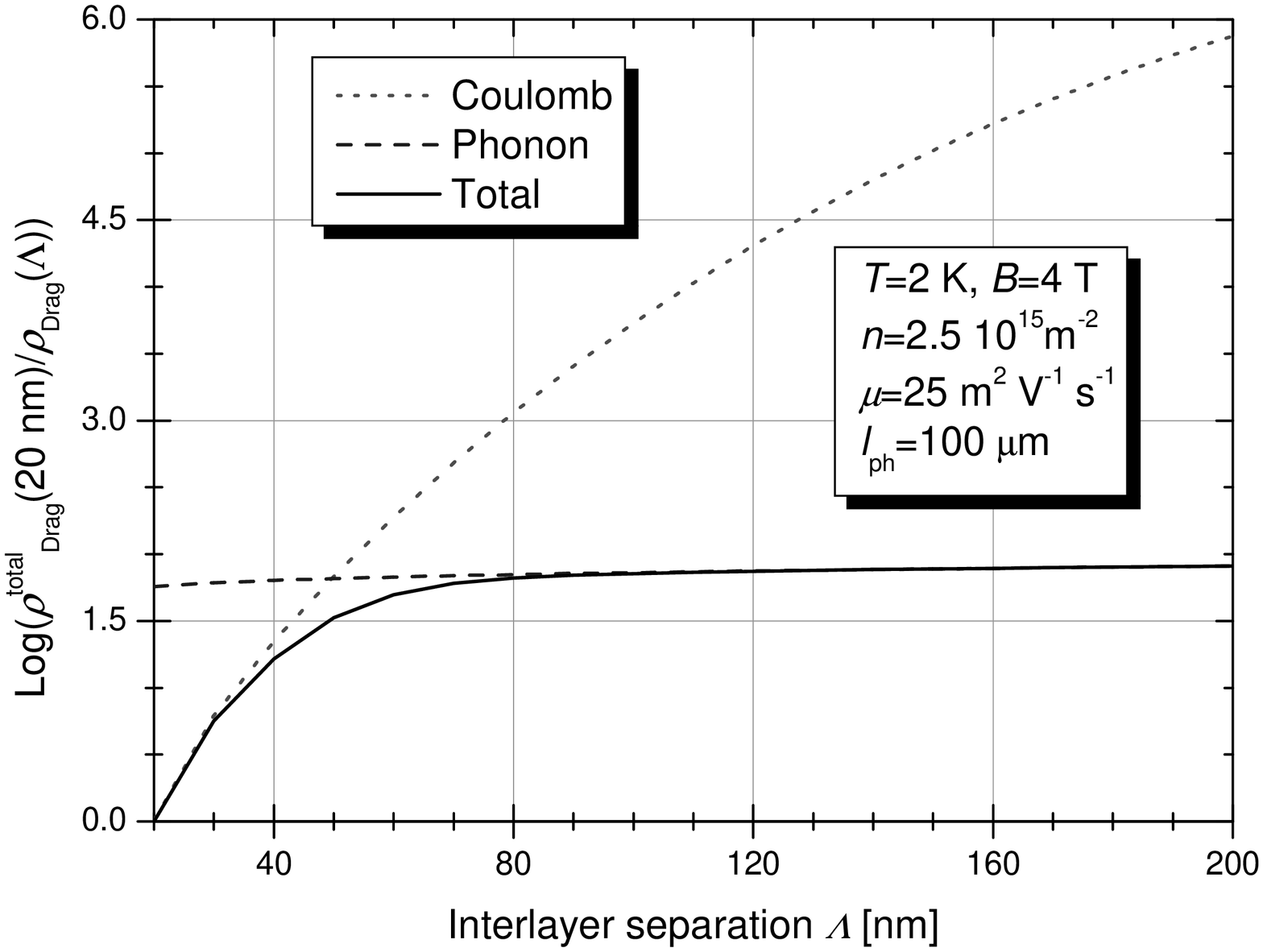}
\caption{Transresistivity as a function of the interlayer separation,
normalized to the value of total transresistivity at $\Lambda=20$ nm. The
contributions of the pure Coulomb and the pure phonon mediated couplings are
shown together with their combined total contribution.}
\label{fg2}
\end{figure}

In Fig.~\ref{fg2} we plot the transresistivity as a function of the
interlayer spacing $\Lambda $ for pure Coulomb $e-e$ interaction, $%
W_{ij}^{C}(q,\omega )$, and for pure PA phonon mediated $e-e$ interaction, $%
W_{ij}^{PA}(q,\omega )$. In addition, we plot the total $\rho _{Drag}$ due
to their combined effect, $W_{ij}(q,\omega )=W_{ij}^{C}(q,\omega
)+W_{ij}^{PA}(q,\omega )$. In all three cases we calculate the screening
function by taking $W_{ij}(q,\omega )=W_{ij}^{C}(q,\omega
)+W_{ij}^{PA}(q,\omega )$. It is seen that the Coulomb mechanism dominates
for $\Lambda <30$ nm with an approximately exponential decrease of $\rho
_{Drag}$ in $\Lambda $ while the phonon mechanism dominates for $\Lambda >80$
nm with an approximately logarithmic decrease. They have comparable
contributions to magnetodrag for the intermediate $\Lambda $, showing the
equal contributions at about $\Lambda =50$ nm.

\begin{figure}[tbp]
\centering
\includegraphics[width=6cm,angle=-90]{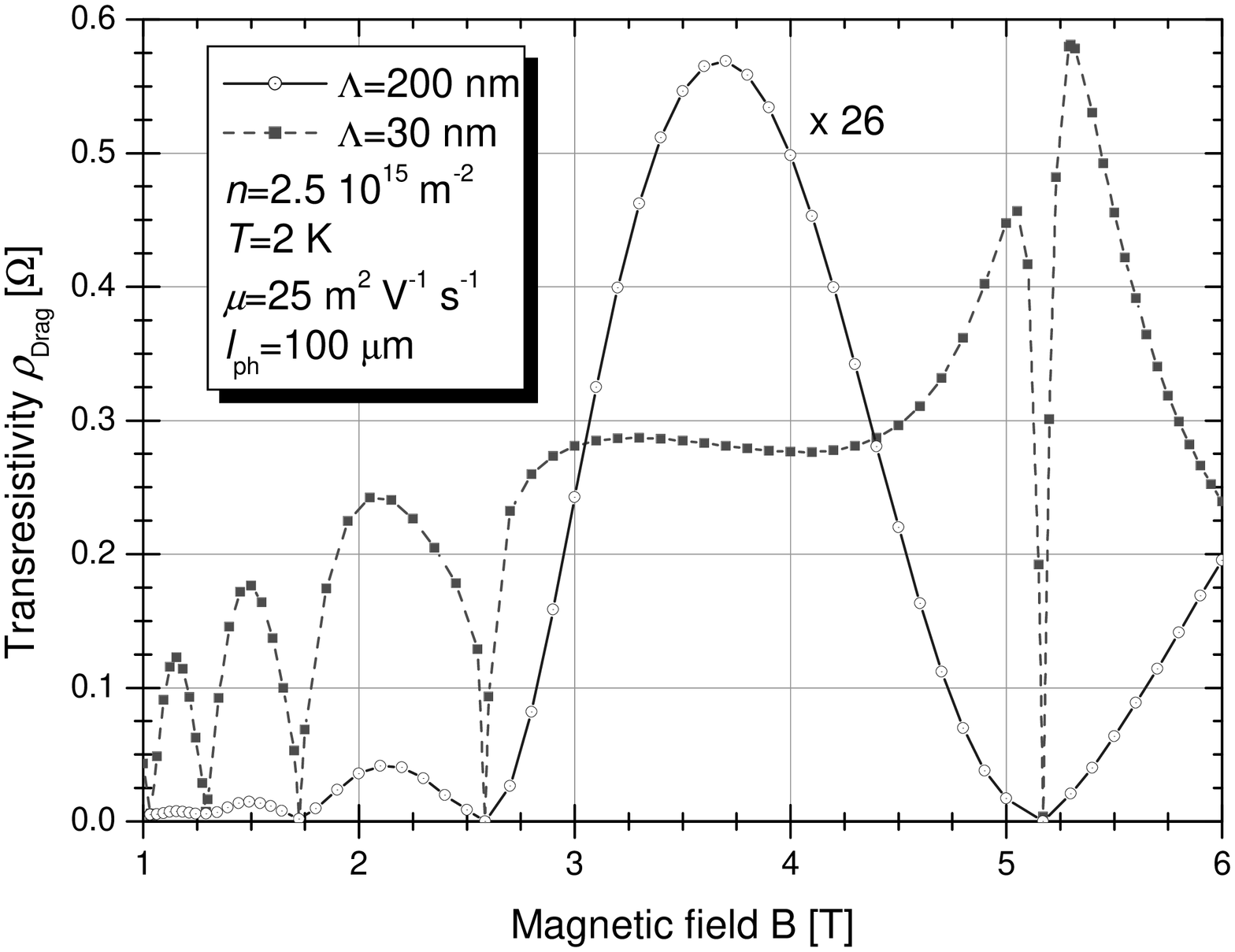}
\caption{Total transresistivity as a function of magnetic field plotted for
the $\Lambda =30$ and $200$ nm interlayer separations. The data for $\Lambda
=200$ nm has been multiplied by a factor of $26$.}
\label{fg3}
\end{figure}

In Fig.~\ref{fg3} we plot the magnetic field dependence of total
transresistivity due to Coulomb and PA\ phonon mediated $e-e$ interaction.
The transresistivity shows Shubnikov-De Haas oscillations as a function of $%
B $ both for the $\Lambda =30$ and $200$ nm spacings. For $\Lambda =200$ nm
we find that $\rho _{Drag}(B)$ has perfectly symmetric oscillations with
strongly increasing amplitudes in $B$. We ascribe this to weak screening at
large interlayer separations. For $\Lambda =30$ nm, in agreement with the
calculations by B\o nsager \textit{et al}. \cite{bons96} for pure Coulomb
interaction, $\rho _{Drag}(B)$ demonstrates a double peak structure around
the filling factor $\nu =3$, caused by strong screening. At low fields
screening is relatively weak and, except for some asymmetry, no dip is
visible for $\nu \geq 4$. However, screening has a flattening effect on the
oscillation amplitudes in different inter-plateau regions for $\Lambda =30$
nm. The phonon contribution to the total magnetodrag for $\Lambda =30$ nm is
about $5 $ to $10\%$ in the second inter-plateau region, depending on $B$.
For $\Lambda =200$ nm the Coulomb contribution to $\rho _{Drag}(B)$ is
negligible. It is seen from Fig.~\ref{fg3} that near $\nu =3$ the total $%
\rho _{Drag}$ is about $0.3$ $\Omega $ for $T=2$ K and $\Lambda =30$ nm, and
this is approximately $13$ times larger than $\rho _{Drag}$ for $\Lambda
=200 $ nm. The magnitude of the calculated $\rho _{Drag}$ and its strong
decrease with increasing $\Lambda$ are in good quantitative agreement with
the experimental findings by Rubel \textit{et} \textit{al}.\ \cite{rubel97e}%
. In this experiment, however, dips are also observed in the middle of
Landau bands for samples with large interlayer separations. The lack of such
dips in our calculations for $\Lambda =200$ nm might be an additional
indication that these dips in the experiments should be attributed rather to
the spin than the screening effect \cite{hill97,joerger2000e,lok}. 
\begin{figure}[tbp]
\centering
\includegraphics[width=11cm,angle=-90]{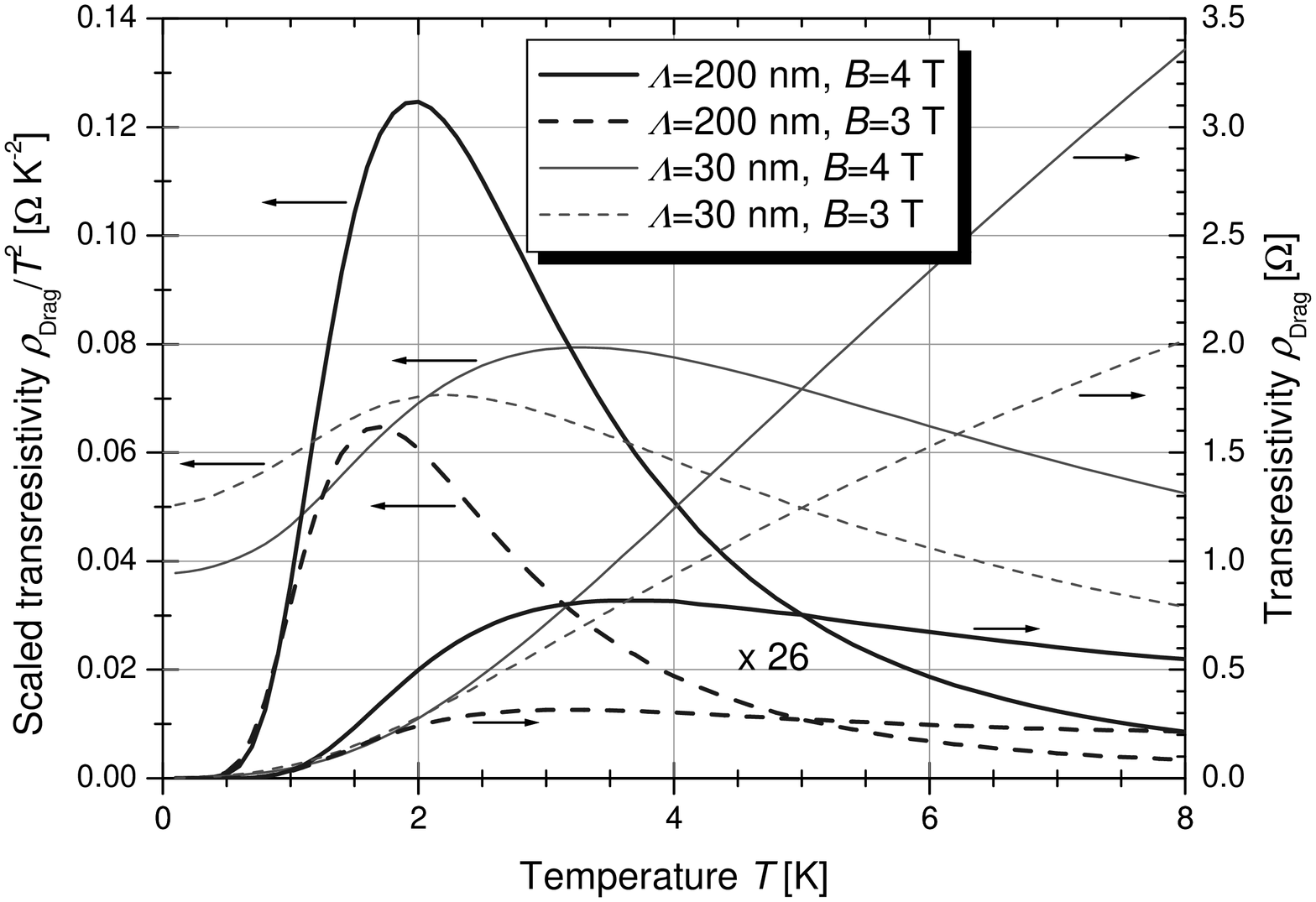}
\caption{Total transresistivity (the left axis) and scaled transresistivity
(the right axis) versus temperature for $B=3$ and $4$ T and for $\Lambda =30$
nm (thin lines) and $200$ nm (thick lines). The data for $\Lambda =200$ nm
has been multiplied by a factor of 26. The electron density is $n=2.5\cdot
10^{15}$ m$^{-2}$, the phonon mean free path $l_{ph}=100$ $\protect\mu$m,
and the mobility $\protect\mu =25$ m$^{-2}$ V$^{-1}$ s$^{-1}$.}
\label{fg4}
\end{figure}

\begin{figure}[tbp]
\centering
\includegraphics[width=6cm,angle=-90]{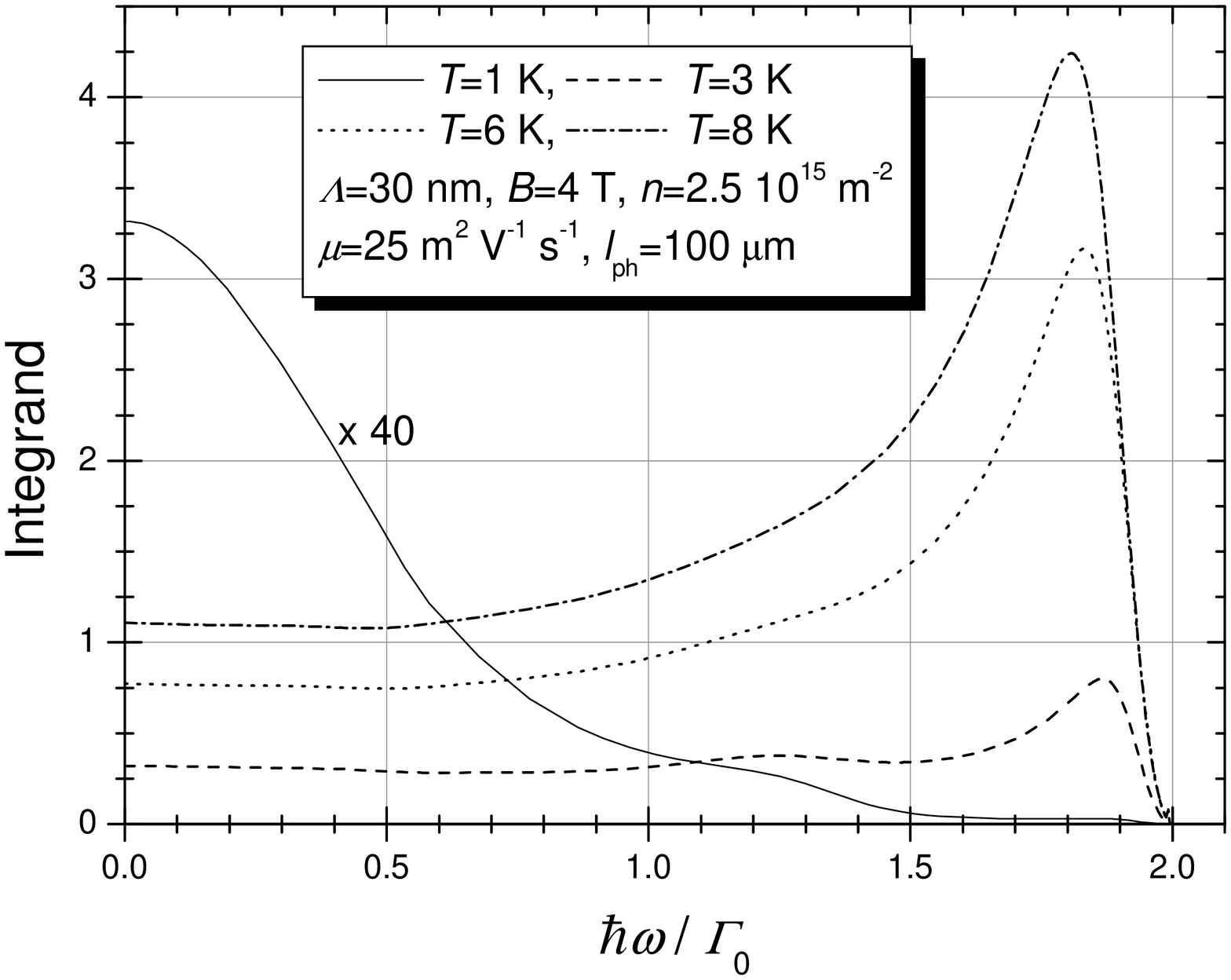}
\caption{Integrand over $\protect\omega $ in Eq.~(\ref{eq2}) after taking
the sum over $\roarrow{q}$ for $\Lambda =30$ nm and for different
temperatures. The data for $T=1$ K has been multiplied by a factor of $40$.}
\label{fg5}
\end{figure}

In Fig.~\ref{fg4} we present the temperature dependence of $\rho _{Drag}$
and $\rho _{Drag}/T^{2}$ for two different values of $B$ and for $\Lambda
=30 $ and $200$ nm in the temperature range from $0.1$ up to $8$ K. Although 
$\rho _{Drag}/T^{2}$ shows a peaked temperature dependence for both $\Lambda
=30$ and $200$ nm, $\rho _{Drag}$ demonstrates completely different behavior
for $\Lambda =30$ and $200$ nm as function of $T$. For $\Lambda =30$ nm the
magnetodrag is mainly due to the Coulomb mechanism and is in agreement with
the calculations by B\o nsager \textit{et al}. \cite{bons96} for pure
Coulomb interaction. At lower temperatures $\rho _{Drag}$ has a subquadratic
temperature dependence and $\rho _{Drag}/T^{2}$ remains finite at low $T$.
The transferred energy $\hbar \omega $ is restricted by temperature, $\hbar
\omega \lesssim T,$ and the transferred momentum is, independently of $\hbar
\omega ,$ restricted by the interlayer spacing, $q\lesssim \Lambda ^{-1}$
(in quantizing magnetic fields $\Lambda ^{-1}$ is always smaller than $\ell
_{B}^{-1}$). This is clearly seen in Fig.~\ref{fg5} 
where we plot the integrand over $\hbar \omega $ in Eq.~(\ref{eq2}) for $%
\Lambda =30$ nm after taking the sum over $\overrightarrow{q}$. Even for
relatively large $T=1$ K the integrand is strongly decreasing function of $%
\omega $. Therefore at low temperatures one can use the small $\omega $ and
the small $q$ limit of the polarization function $\chi (q,\omega )$. In this
limit both $\text{Im}~\chi (q,\omega )$ and $\text{Im}~\varepsilon (q,\omega
)$ are linearly vanishing functions with $\omega $ while Re~$\varepsilon
(q,\omega )$ has a large maximum, determined by the electronic
compressibility. By replacing the upper limit of the integration over $\hbar
\omega $ in Eq.~(\ref{eq2}) with $T$ and taking the static limit of the
screening function out of the integration, one can obtain (cf. Fig.~\ref{fg4}%
) 
\begin{equation}
\rho _{Drag}\propto T^{2}\int_{0}^{T}\frac{d\hbar \omega }{T}\frac{\left( 
\text{Im}\chi \left( \hbar \omega \right) /T\right) ^{2}}{\sinh ^{2}\left(
\hbar \omega /2T\right) }\propto T^{2}.  \label{eq19}
\end{equation}
At high temperatures all $\hbar \omega <2\Gamma _{0}$ contribute to the
integral in Eq.~(\ref{eq2}). One can see from Fig.~\ref{fg5} that for $T=3$, 
$6$, and $8$ K the integrand decreases very slowly in the region of small $%
\omega $. However, at energies close to $\hbar \omega ^{-}\left( 0\right) $,
when the first magnetoplasmon mode is developed (cf. Fig.~\ref{fg6}), the
integrand starts to increase slowly. Further, at energies close to $\hbar
\omega ^{+}\left( 0\right) $, when the second magnetoplasmon mode is
developed (cf. Fig.~\ref{fg6}), the integrand starts to increase strongly.
The integrand reaches its well pronounced maximum, with an increasing height
in $T$, just before it drops to zero at the upper edge of the Landau band $%
\hbar \omega =2\Gamma _{0}$. This behavior of the integrand results in the
peaked temperature dependence of $\rho _{Drag}/T^{2}$ shown in Fig.~\ref{fg4}
with an approximately linear temperature dependence of $\rho _{Drag}\propto
T $ at high temperatures, observed experimentally by Hill \textit{et al.} 
\cite{hill97}.

\begin{figure}[tbp]
\centering
\includegraphics[width=6cm,angle=-90]{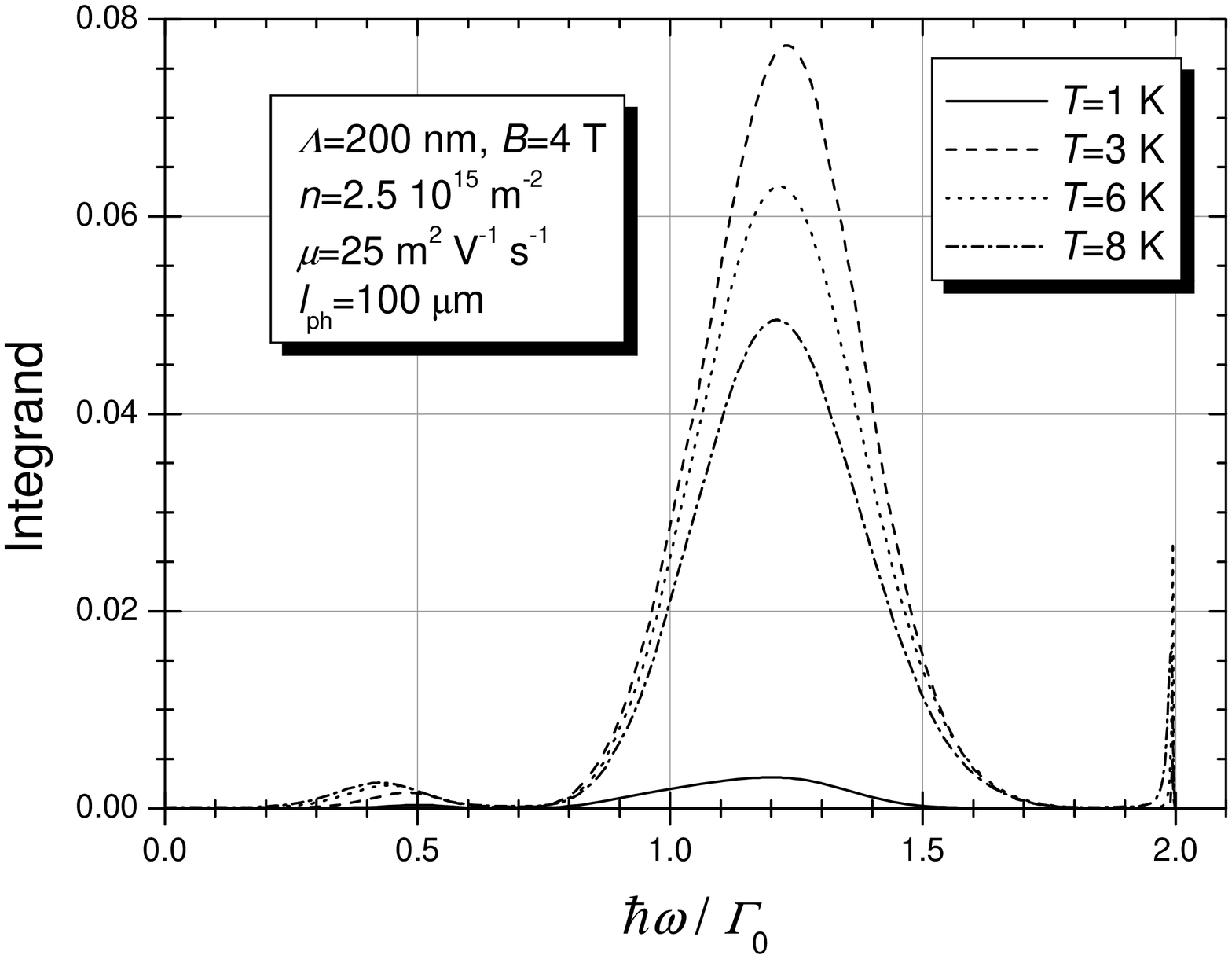}
\caption{Integrand over $\omega $ in Eq.~(\ref{eq2}) after taking
the sum over $\roarrow{q}$ for $\Lambda =200$ nm and for different
temperatures.}
\label{fg7}
\end{figure}

The situation is completely changed for $\Lambda =200$ nm where the
magnetodrag is mainly due to phonon exchange. At low temperatures $\rho
_{Drag}$ has stronger dependence on $T$ than it has at the small interlayer
separation. At high temperatures $\rho _{Drag}$ decreases slowly with
increasing $T$. This results in the very sharp peak of $\rho _{Drag}/T^{2}$
as a function of $T$ for $\Lambda =200$ nm (Fig.~\ref{fg4}). This strikingly
different behavior of $\rho _{Drag}$ as a function of $T$ for different $%
\Lambda $ is a consequence of the fact that the main contribution to the
phonon mediated drag comes from $q\approx \omega /s$ for which effective
unscreened $e-e$ interaction diverges in the limit of infinite phonon
lifetime. Now the integrand over $\omega $ in Eq.~(\ref{eq2}) is
proportional to $g_{l}(t)\equiv t\left( L_{l}(t)\right) ^{2}e^{-t}$ with $%
t=\left( \omega /s\right) ^{2}\ell _{B}^{2}/2$ and in the whole temperature
range the main contribution to phonon magnetodrag makes $t=t_{l}\sim 1$ ($%
t_{l}$ are the zeros of the derivative of $g_{l}(t)$)$,$ i.e. the finite $%
\hbar \omega \approx \hbar sq\sim \hbar s/\ell _{B}\sim \Gamma _{0}$. The
small $\omega $ and the small $q$ approximation is not valid for $%
\varepsilon (q,\omega )$ and it is rapidly decreasing function with $\omega $
and the screening effect is weak in the phonon magnetodrag (cf. the curves
corresponding to $\Lambda =30$ and $200$ nm in Fig.~\ref{fg3}). We obtain
the following formula as an approximation to describe the temperature
dependence of the transresistivity in this regime 
\begin{equation}
\rho _{Drag}\sim \sum_{l}{\frac{1}{\nu ^{2}}}\frac{t_{l}}{\vartheta }\frac{%
\left| W_{12}(t_{l},y_{l})\right| ^{2}}{\left| \varepsilon
(t_{l},y_{l})\right| ^{2}}\frac{\left( \text{Im}\chi (t_{l},y_{l})\right)
^{2}}{\sinh ^{2}(y_{l}/2\vartheta )}
\end{equation}
where $\vartheta \equiv T/\Gamma _{0}$ and $y\equiv \hbar \omega /\Gamma
_{0} $. For $B=4$ T we have $\nu =2$ and the main contribution to the sum
makes the outermost filled Landau level with $l=1$ and $t_{1}=2+\sqrt{3}$.
In GaAs $2ms^{2}\approx 0.23$ K and $\Gamma _{0}\approx 6.5$ K for $\mu =25$
m$^{2}$V$^{-1}$s$^{-1}$, so we have $t\approx \left( 1.5y\right) ^{2}$. This
gives $\hbar \omega _{1}/2\approx 4.2$ K as a crossover temperature of the
temperature dependence of $\rho _{Drag}(T)$. At temperatures below $\hbar
\omega _{1}/2,$ $\sinh ^{2}(\hbar \omega _{1}/2T)$ increases strongly with $%
T $ and this determines mainly the behavior of $\rho _{Drag}(T)$. It is seen
from Fig.~\ref{fg4} that for $\Lambda =200$ nm $\rho _{Drag}/T^{2}$ tends to
zero at low $T$ in contrast to its finite value for $\Lambda =30$ nm. At
temperatures above $\hbar \omega _{1}/2,$ $\sinh ^{2}(\hbar \omega
_{1}/2T)\propto T^{2}$, and the screening function $\varepsilon
(q_{1},\omega _{1})$ for such large arguments is small and increasing in $T$%
. This behavior together with the $T^{-1}$ prefactor in Eq.~(\ref{eq2})
determines the temperature dependence of $\rho _{Drag}$ at high $T$. One can
see, however, from Fig.~\ref{fg4} that the actual peak position of $\rho
_{Drag}(T)$ is shifted to the left from $\hbar \omega _{1}/2$. For $\Lambda
=200$ nm the integrand over $\omega $ in Eq.~(\ref{eq2}) after taking the
sum over $\overrightarrow{q}$ manifests itself to three strongly pronounced
peaks near $y=0.45$, $1.25$, and $2$, and their positions are relatively
stable with respect to temperature (see Fig.~\ref{fg7}). 
Notice, as against the situation of the Coulomb magnetodrag, even for small $%
T=1$ K the integrand is approximately zero in the region of small $\omega $
and has strongly pronounced peak at large $\omega $. The main highest peak
of the integrand near $y=1.25$, as we already discussed, is related to the
maximum of $g_{1}(t)$ at $t_{1}^{+}=2+\sqrt{3}$ corresponding to $%
y=y_{1}^{+}=1.29.$ The left small peak of the integrand near $y=0.45$ is
related to another maximum of the function $g_{1}(t)$ at $t_{1}^{-}=2-\sqrt{3%
}$ ($y_{1}^{-}=0.35$) and also to the maximum of the function $g_{0}(t)$ at $%
t_{0}=1$ ($y_{0}=0.67$). This second peak contribution to $\rho _{Drag}(T)$
shifts the actual crossover temperature of $\rho _{Drag}(T)$ to the left so
that $T_{peak}\approx 3.6$ K (Fig.~\ref{fg4}). One can see, however, that
the positions of the first two peaks of the integrand is not far from the
magnetoplasmon energies $\hbar \omega ^{\pm }\left( 0\right) .$ Therefore
the magnetoplasmons contribute also to the formation of these peaks to some
extend. Nevertheless, it is clear from the very weak dependence of the total 
$\rho _{Drag}$ on $\Lambda $ (Fig.~\ref{fg2}) that the magnetoplasmon
contribution to $\rho _{Drag}$ for $\Lambda =200$\ nm is a small correction
to the phonon contribution. The third peak of the integrand in the very
close vicinity of the upper edge of the Landau band is a pure magnetoplasmon
effect (cf. Fig.~\ref{fg6} and Fig.~\ref{fg7}). Immediately below the upper
edge of Landau band the magnetoplasmon damping is very small and this
results in the very sharp peak of the integrand with a relatively small
contribution to $\rho _{Drag}(T)$. Thus, at high temperatures we obtain that
for $\Lambda =200$ nm $\rho _{Drag}$ decreases slowly with $T$ (Fig.~\ref
{fg4}). This extraordinary temperature dependence of $\rho _{Drag}$ is
mainly due to the weak screening effect at large interlayer separations and
the screening is weak because the phonons with large energies and momenta
realize $e-e$ interaction between remote layers. Notice that the $T^{-1}$
dependence of $\rho _{Drag}(T)$ for the pure Coulomb interlayer interaction
in one of the typical scattering regions was also reported by Khaetskii and
Nazarov \cite{khaetskii}. As far as we know at present no magnetodrag
measurements are available in the above regime and the experimental test of
the temperature dependence of phonon magnetodrag is critical.

\section{Summary and conclusions}

We have calculated the transresistivity between spatially separated electron
layers in a perpendicular magnetic fields. We take into account both direct
electrostatic Coulomb and effective phonon mediated $e-e$ interaction. For
this system we calculate the dispersion relation of the intra-Landau level
magnetoplasmons within the random phase approximation at the finite
temperatures and distinguish the magnetoplasmon contribution to the
magnetodrag. We find the strikingly different magnetic field and temperature
dependence of the transresistivity for the small and large interlayer
separations. When $\Lambda =30$ nm, the transresistivity shows a slight dip
as a function of $B$ in the middle of the second spin-degenerated Landau
band with the total $\rho _{Drag}$ about $0.3$ $\Omega $ (the phonon mean
free path is $100$ $\mu $m, the mobility $\mu =25$ m$^{2}$ V$^{-1}$ s$^{-1}$%
, $T=2$ K, and $n=2.5$ $\cdot 10^{15}$ m$^{-2}$), which, depending on $B$,
includes about $5$ to $10\%$ of the phonon contribution due to piezoelectric
interaction. When $\Lambda =200$ nm, the Coulomb contribution to magnetodrag
is negligible. And, $\rho _{Drag}$ shows no dip and the total drag around
filling factor $\nu =3$ is about $13$ times less than the value for $\Lambda
=30$ nm. This is in good agreement with the experimental findings by Rubel 
\textit{et} \textit{al}.\ \cite{rubel97e}. The scaled transresistivity $\rho
_{Drag}/T^{2}$ has a peaked temperature dependence for both $\Lambda =30$
and $200$ nm. At low $T$, $\rho _{Drag}/T^{2}$ remains finite for $\Lambda
=30$ nm, while for $\Lambda =200$ nm it tends to zero. At high $T$, $\rho
_{Drag}$ is approximately linear in $T$ for $\Lambda =30$ nm while for $%
\Lambda =200$ nm it decreases slowly with $T$. Therefore, the peak of $\rho
_{Drag}/T^{2}$ as function of $T$ is very sharp for the large separation $%
\Lambda =200$ nm. We ascribe this to the weak screening effect at large
interlayer separations where the phonon mechanism dominates with the large
transferred energy and momentum.

\begin{acknowledgments}
This work was supported by Chonnam National University under a grant in the
year of 2002. We are grateful to R. Gerhardts, W. Dietsche, and 
K. von Klitzing for useful discussions. SMB acknowledges the support and 
hospitality in Max-Planck-Institute for Solid State Researches where 
this work was initiated.
\end{acknowledgments}

\end{document}